\documentclass[twocolumn,epjc3]{svjour3}

\RequirePackage[T1]{fontenc}

\smartqed  % flush right qed marks, e.g. at end of proof

\RequirePackage{graphicx}
\RequirePackage{mathptmx}      % use Times fonts if available on your TeX system
\RequirePackage{flushend}
\RequirePackage[numbers,sort&compress]{natbib}
\RequirePackage[colorlinks,citecolor=blue,urlcolor=blue,linkcolor=blue]{hyperref}
\RequirePackage{amsmath}
\usepackage{threeparttable}
\usepackage{lineno}
\journalname{Eur. Phys. J. C}

\begin{document}

\title{Performance study of particle identification at the CEPC using TPC  $dE/dx$  information}
%\title{Insert your title here\thanksref{t1}}


\author{F. An\thanksref{e1,addr1,addr3}
       \and
       S. Prell\thanksref{addr3}
       \and
       C. Chen\thanksref{addr3}
       \and
       J. Cochran\thanksref{addr3}
        \and
        X. Lou\thanksref{addr1,addr2,addr4}
        \and
        M. Ruan\thanksref{e2,addr1}
}

%\thankstext[$\star$]{t1}{Thanks to the title}
\thankstext{e1}{e-mail: anff@ihep.ac.cn}
\thankstext{e2}{e-mail: manqi.ruan@ihep.ac.cn}

\institute{Institute of High Energy Physics, Chinese Academy of Science, Beijing, China\label{addr1}
   \and
         Department of Physics and Astronomy, Iowa State University, Ames IA, USA\label{addr3}
          \and
          Physics Department, University of Texas at Dallas, Richardson TX, USA\label{addr2}
                \and
        University of Chinese Academy of Science (UCAS), Beijing, China\label{addr4}
}

\date{Received: date / Accepted: date}

\maketitle

\begin{abstract}
The kaon identification is crucial for the flavor physics, and also benefits the flavor and charge reconstruction of the jets.
We explore the particle identification  capability 
for tracks with momenta ranging from  2-20 GeV/c
using  the  $dE/dx$  measurements
in the Time Projection Chamber  at the future Circular Electron-Positron Collider.
Based on Monte Carlo simulation, we anticipate
  that  an average   $3.2~\sigma$ ($2.6~\sigma$) $K/\pi$ separation  can be achieved 
based on $dE/dx$ information
for an optimistic (conservative) extrapolation of the simulated performance to the final system.
Time-of-flight (TOF) information from the Electromagnetic Calorimeter
can provide $K/\pi$ separation around 1 GeV/c and reduce the $K/p$ mis-identification rate.
By combining the $dE/dx$  and TOF  information, we estimate that in the optimistic scenario 
a kaon  selection in 
inclusive hadronic $Z$ decays with both the average efficiency and purity approaching 95\% can be achieved.

\end{abstract}

\section{Introduction}
\label{sec:introduction}

The project of building a Circular Electron-Positron Collider (CEPC) \cite{cepc_precdr} in China has been proposed.
The CEPC will operate as a Higgs boson factory or a $Z$ boson factory at center-of-mass energies of $\sqrt{s}\sim$ 
240 or 91 GeV, respectively. During the lifetime of the CEPC,  one million Higgs bosons are expected to be produced,
allowing  precision measurements of the Higgs boson properties~\cite{manqi}. In addition, ten billion $Z$  bosons will 
be delivered at the $Z$ pole promising refined measurements of electroweak and heavy flavor physics~\cite{ew_physics}.

A Time Projection Chamber (TPC) has been proposed as a candidate charged particle tracking device for the CEPC 
detector. TPCs have been operated successfully in $e^+e^-$ and hadron collider experiments and even in fixed-target 
experiments, such as the ALEPH~\cite{aleph1} and ALICE~\cite{alice1} experiments at CERN, 
 the HISS experiment at BEVALEC \cite{hiss}, etc. A TPC provides precise momentum and position measurements, 
a low material budget, and good particle identification (PID) over a wide range of momentum. The PID information is 
based on $dE/dx$ measurements in the TPC, where $dE/dx$  is defined as the energy deposit per unit path length. 
There are several ongoing R\&D efforts about the TPC proposal, 
such as exploring novel technologies of the  GEM-Micromegas \cite{zhangyl} or GEM \cite{gem_ilc} readout detectors, 
the voxel occupancy  in the TPC in decays at the $Z$ pole \cite{zhaom}, etc.
Compared to previous TPCs, we expect an improved performance of the proposed TPC at the 
CEPC detector as a result of the increased number of readout channels and recent developments in readout electronics.

In this paper, the $dE/dx$ performance of the CEPC TPC is investigated based on Monte Carlo (MC) 
simulation. 
PID will play an important role in measurements of the bottom ($b$) and charm ($c$) hadron
decays in heavy flavor physics. It can  also be exploited to enhance the flavor 
tagging of the $b/c$-jets in Higgs and precision electroweak measurements.
We study the PID of kaons, pions and protons in hadronic decays at the $Z$ pole, demonstrating
that an effective kaon selection can be achieved by combining the $dE/dx$ measurements of the TPC with the time-of-flight 
(TOF) information provided by the Electromagnetic Calorimeter (ECAL) at the CEPC detector.
      
%The dE/dx measurements played an important role in many heavy flavor 
%measurements in the LEP experiments \cite{opal_1,opal_2,opal_3,opal_4,opal_5,aleph_1}.
%like the average polarization of b baryons~\cite{opal_1},
%the asymmetry in the population of particles in the regions between jets~\cite{opal_2},
%the production of $D^{*\pm}$ or $J$/$\psi$ in $Z$ decays~\cite{opal_3,opal_4},
%and heavy flavor production in $Z$ decays~\cite{aleph_1}.
%Its use in more recent  heavy flavor measurements from OPAL is also shown 
%in~\cite{opal_5}.
%At the CEPC, two orders of magnitude more $Z$ bosons will be produced compared to that at the LEP.

% in $Z$ decays, kaon and pion larger than 2 GeV/c can be separated using  $dE/dx$  
%with an average PID efficiency of about 86\% and mis-identification probability of 1\%.
%The efficiency can become 94\% if we enlarge the TPC outer radius.

%In a certain gas, the  $dE/dx$  of particles 
%should depend only on the relativistic velocity $\beta\gamma$
%regardless of the particle type. Therefore by measuring the momentum
%and  $dE/dx$ , the mass and thus the type of the particle can be decided.
%The separation power between different particle species
 %depends both on the  $dE/dx$  difference and the resolution.
%The resolution is determined by the TPC configuration, the imperfect calibration procedure
% about the gas instability, 
%track length and so on, and the electronic noise.

The paper is organized as follows. In section~\ref{sec:dedx_mean} we present the configuration of the CEPC TPC 
 and the energy loss measurement of traversing charged particles. Section~\ref{sec:dedx_res} describes 
the key factors influencing the  resolution of the $dE/dx$ measurement   and provides an estimate of the PID 
performance at the CEPC. In section~\ref{sec:conclusion} a brief conclusion is given.

\section{Energy deposit  in TPC}
\label{sec:dedx_mean}

The TPC concept was introduced in Ref. \cite{1tpc}.
%firstly proposed by David Nygren in 1975
%The TPC  is a gaseous detector which provides 
%3-D track information and  $dE/dx$  measurements for PID.
%To illustrate the typical TPC structure, we depict
%the ALICE TPC in Fig.~\ref{fig:tpc}  as an example.
 %The main structure has been introduced in Ref.~\cite{cepc_precdr}.
A TPC consists essentially of a wireless drift volume situated between parallel axial electric 
and magnetic fields, where the electric field is set up between a central cathode plate and the 
end plates. When a charged particle traverses the gas-filled drift volume, it generates
electron-ion pairs by collisions with the gas molecules. The electrons drift towards the end plates,
where the charges are amplified and collected.

%{\red The figure of ALICE TPC is removed because it is not the CEPC TPC and may mis-lead readers.}
%\begin{figure}[htp]
%  \centering
%  \includegraphics[width=0.55\linewidth]{pic/tpc.eps}
%  \caption{Sketch of the ALICE TPC.}
%  \label{fig:tpc}
%\end{figure}

The default design of the TPC at the CEPC detector  can be found in Ref. \cite{cepc_precdr}.
It is a cylindrical detector that is 4.7 m long with an  inner and outer radii of 0.325 m and 1.8 m, respectively.
The candidate gas is an argon-based gas composite (93\% Ar+5\% CH$_{4}$+2\% CO$_{2}$)
 held at atmospheric pressure and room temperature.
A solenoid provides a magnetic field of 3 T along the beam direction.
In the endcaps, Micromegas  \cite{micromegas} detector modules with pad size of 6 mm along the 
radial direction (height) and 1 mm along the azimuthal direction (width) are arranged in 222 concentrical 
rings.

%The mean energy deposit  of particles in matter, denoted as $I$, is described by 
% the Bethe equation~\cite{pdg}:
%\begin{equation}
%  I =
%%4\pi N_{\alpha}r_{e}^{2}m_{e}c^{2}z^{2}
%  0.3071z^{2}
%  (\frac{Z}{A})
%  (\frac{1}{\beta^2})
%  \left[\frac{1}{2} \ln\left(\frac{2m_{e}c^2\beta^2\gamma^2E_{cut}}{W^{2}}\right)-\beta^2-\frac{\delta}{2}\right],
%  \label{eq:dedx_mean}
%\end{equation}
%
% \begin{equation}  
%    \delta=
%    \left\{  
%                 \begin{array}{lr}  
%           0,
%               &   x=\log_{10}(\beta\gamma)<x_{0} 
%               \vspace{0.15cm}  \\
%                  
%             2\ln(x)-\bar{C}+a(x_{1}-x)^{k},
%               &   x_{0} \le x \le x_{1}
%                 \vspace{0.15cm}  \\
%                  
%             2\ln(x)-\bar{C},
%               &   x \ge x_{1}
%               
%                 \end{array}  
%    \right.  
%    \label{eq:delta}
%    \end{equation}  
%where $z$ is the charge of the projectile, $Z$ and $A$ the atomic number and mass,
%and $W$ the mean excitation energy in a given gas mixture.
%$E_{cut}$ represents the maximum energy transfer in a single collision.
%$\delta$ is a function of $\beta\gamma$ to correct the density effect caused by the polarization
%of the medium, which we compute using Sternheimer's parameterization~\cite{Fermi:1940,R.M. Sternheimer:1984}.
%Except $E_{cut}$,
%{ which is determined  by fits to our MC results,
%(as seen from Ref.~\cite{Va'Vra:1999}),}
% the values of all the other parameters in Eq.~(\ref{eq:dedx_mean}-\ref{eq:delta}) 
%are taken from Ref.~\cite{R.M. Sternheimer:1984}.

In the MC simulation, the description of the detector geometry, 
material and  the ionization process are  implemented using GEANT4 \cite{geant4}.
Single-particle events are generated using ParticleGun.
Collision events of the Standard Model processes are produced with the event generator WHIZARD \cite{whizard}.
The  $dE/dx$ measurement by each pad is defined as the energy deposit 
divided by the track length in the corresponding drift volume, both of which are provided by GEANT4.
Typically, the $dE/dx$ measurements of a track follow a Landau distribution
with a large tail caused by high-energy $\delta$-electrons.
We estimate a representative average $dE/dx$ for a track, denoted as $I$,
by using the common ``truncated mean'' method \cite{truncate}.
We calculate $I$ as the mean of the lowest 90\% of  the $dE/dx$  values associated with the track,
where the truncation ratio of 90\%  is determined to yield the optimal $dE/dx$ resolution. 
The distribution of the truncated mean $I$ can be  well described  by a Gaussian 
function with a width denoted as $\sigma_I$. Unless explicitly stated, the $dE/dx$ resolution in the 
paper refers to the ratio $\sigma_{I}/I$.

%Single track events  are simulated with  momentum $p$ between 1 and 100 GeV/c 
%and the  incident angle $\theta=45^{o}$ relative to the beam direction.
%Such a direction ensures that the  projectile can traverse the full radius
%and almost have the best resolution.

\begin{figure*}
  \centering
  \includegraphics[width=0.495\linewidth]{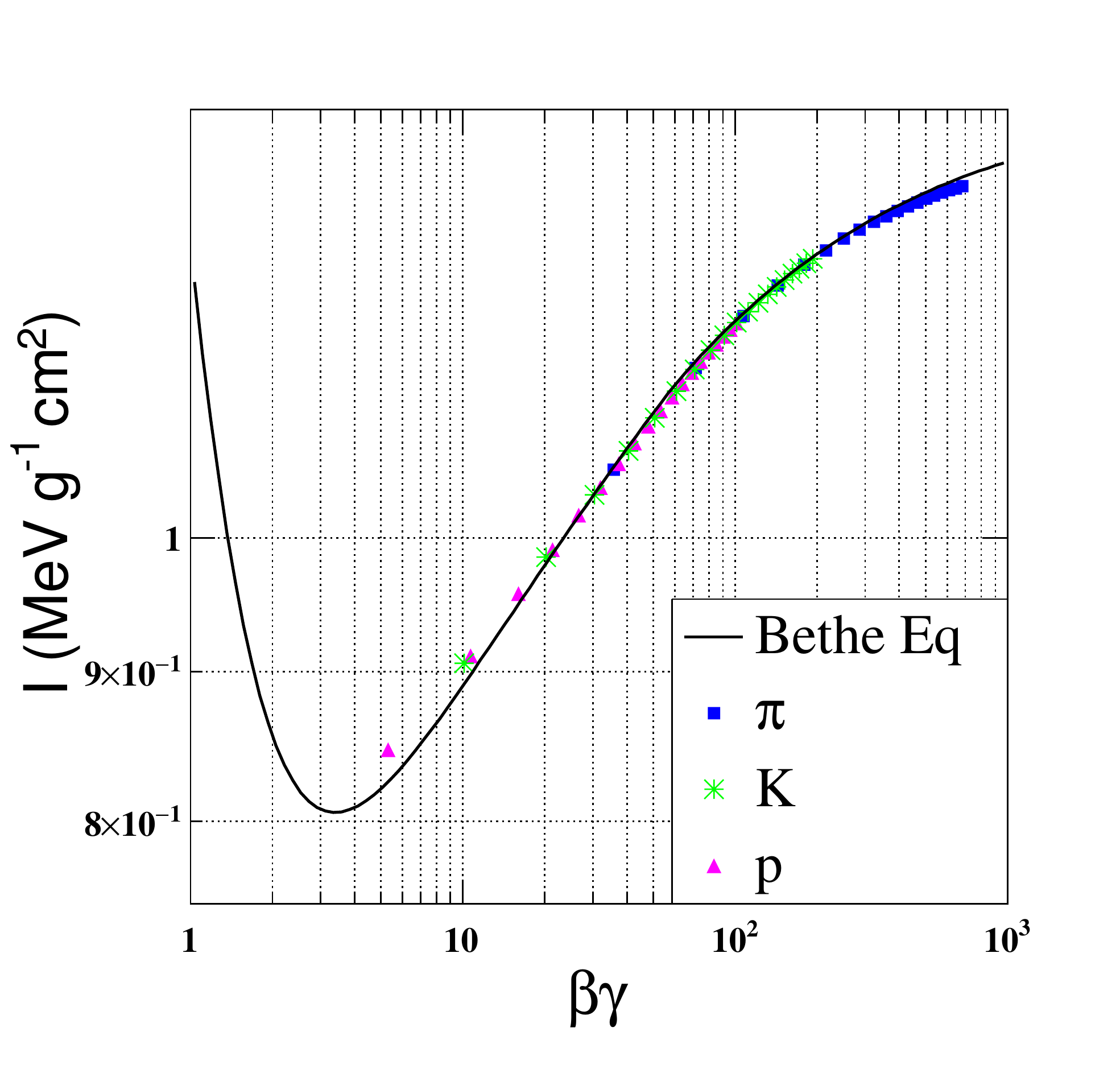} \hfill
    \includegraphics[width=0.495\linewidth]{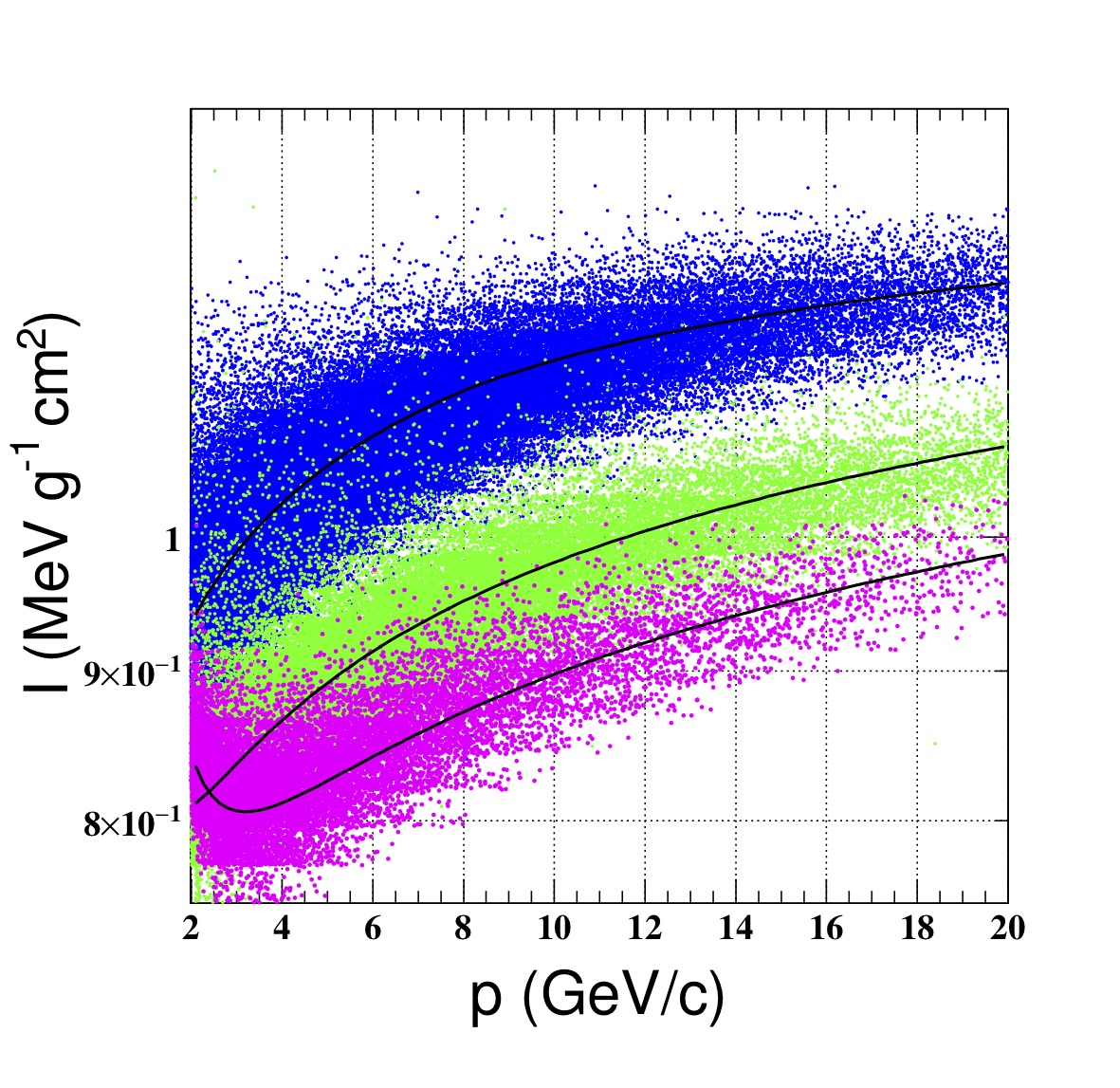}
  \caption{The dependence of the truncated mean $I$ of the track  $dE/dx$,   as a function of
    $\beta\gamma$  (left) and  $p$  (right) for charged particles 
   traversing   the  TPC of the CEPC detector.
In the left plot  the dots represent the MC result of single-particle events
with the theoretical prediction by the Bethe equation~\cite{pdg} overlaid.
In the right plot the dots are from simulation of  $e^{+}e^{-}\to Z\to q\bar{q}$ events.
} 
  \label{fig:dedxMean2bg}
\end{figure*}

For a particle with momentum $p$ and mass $m$,  the MC simulation of
the dependence of $I$  as a function of $\beta\gamma=p/(m c)$ 
is shown in the left plot of Fig.~\ref{fig:dedxMean2bg}.
 Herein we use single-particle events requiring  $\theta=45^{o}$ 
so that the tracks traverse the full TPC radius, where $\theta$ is defined as the polar angle
of the tracks with respect to the beam direction.
The simulated $I$ dependence agrees well  with the theoretical prediction by the Bethe equation \cite{pdg}.
The values of all the parameters  in the Bethe equation are taken from Ref.~\cite{R.M. Sternheimer:1984}
except for the normalization scale factor and the   maximum energy transfer $W_{\rm{max}}$,
which is free in the fit to the $I$ distribution following the procedure in  Ref.~\cite{Va'Vra:1999}.
%\footnote{
%The Bethe equation is taken from the PDG~\cite{pdg} with the parameter values taken from Ref.~\cite{R.M. Sternheimer:1984}.
%The maximum energy transfer $W_{\rm{max}}$
% is determined  by fits to the MC results as 851 eV, 
%just as in Ref.~\cite{Va'Vra:1999}.
%}.
%It is fitted with Eq.~(\ref{eq:delta}) and $E_{cut}=851$ eV is determined.
%To investigate the  $dE/dx$  behavior, we start with the simplest case of single track events.
%By default, we set the incident angle $\theta=45^{o}$ relative to the beam direction,
% which is close to the angle with the best resolution and 
%can make the projectile  traverse the full radius.
%The ionizing energy $I$ at low $\beta\gamma$ decreases with $1/\beta^{2}$, 
%reaches a minimum around $\beta\gamma=3.6$, and continues with a
%ogarithmic rise (``relativistic rise region'') until it saturates (``Fermi plateau'').
In the right plot of Fig.~\ref{fig:dedxMean2bg}, the scatter plot of $I$  versus $p$ is presented 
 using a simulated sample of $e^{+}e^{-}\to Z\to q\bar{q}$ events.
At the CEPC, the majority of the particles traversing the TPC have a momentum above 1 GeV/c 
and reside in the relativistic rise region, where TOF measurements can not effectively distinguish 
between the different particles types.
%where the difference in $I$ between particle types is not so large as that in the low momentum region.
Therefore, improving the $dE/dx$ resolution will directly benefit the PID performance.

\section{Resolution of energy deposit in TPC}
\label{sec:dedx_res}

In an ideal case, the $dE/dx$ resolution  for a given track depends on the
number of the $dE/dx$ measurements along the particle trajectory and the number of the ionizing
electrons per measurement.  We name the induced resolution from these factors ``intrinsic $dE/dx$ resolution''.
The  resolution in real experiments, named as ``actual $dE/dx$ resolution'', 
will be deteriorated by the detector effects arising
in the processes of electron drift, signal amplification and readout in TPC.
A detailed study of those effects is beyond the scope of the paper. 
In this paper, we study the intrinsic $dE/dx$ resolution of the CEPC TPC using MC simulation,
and estimate the degradation of the actual resolution 
by comparing the MC-based results with the experimental measurements of previous TPCs.

\subsection{Parameterization of the intrinsic $dE/dx$ resolution}
\label{sec:res_mc}

\begin{figure*}
\centering
  \includegraphics[width=0.32\linewidth]{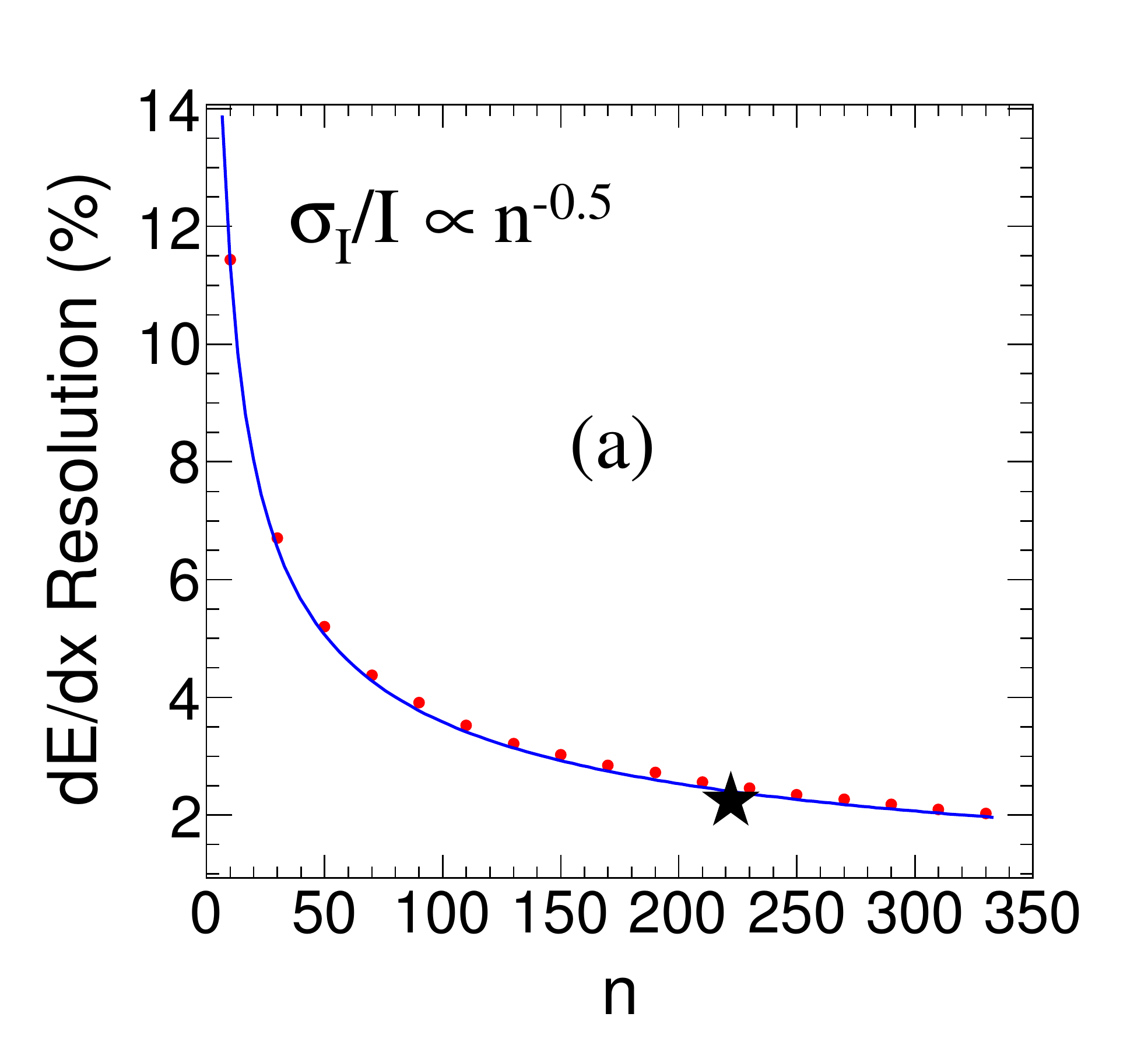}
  \hfill
 \includegraphics[width=0.32\linewidth]{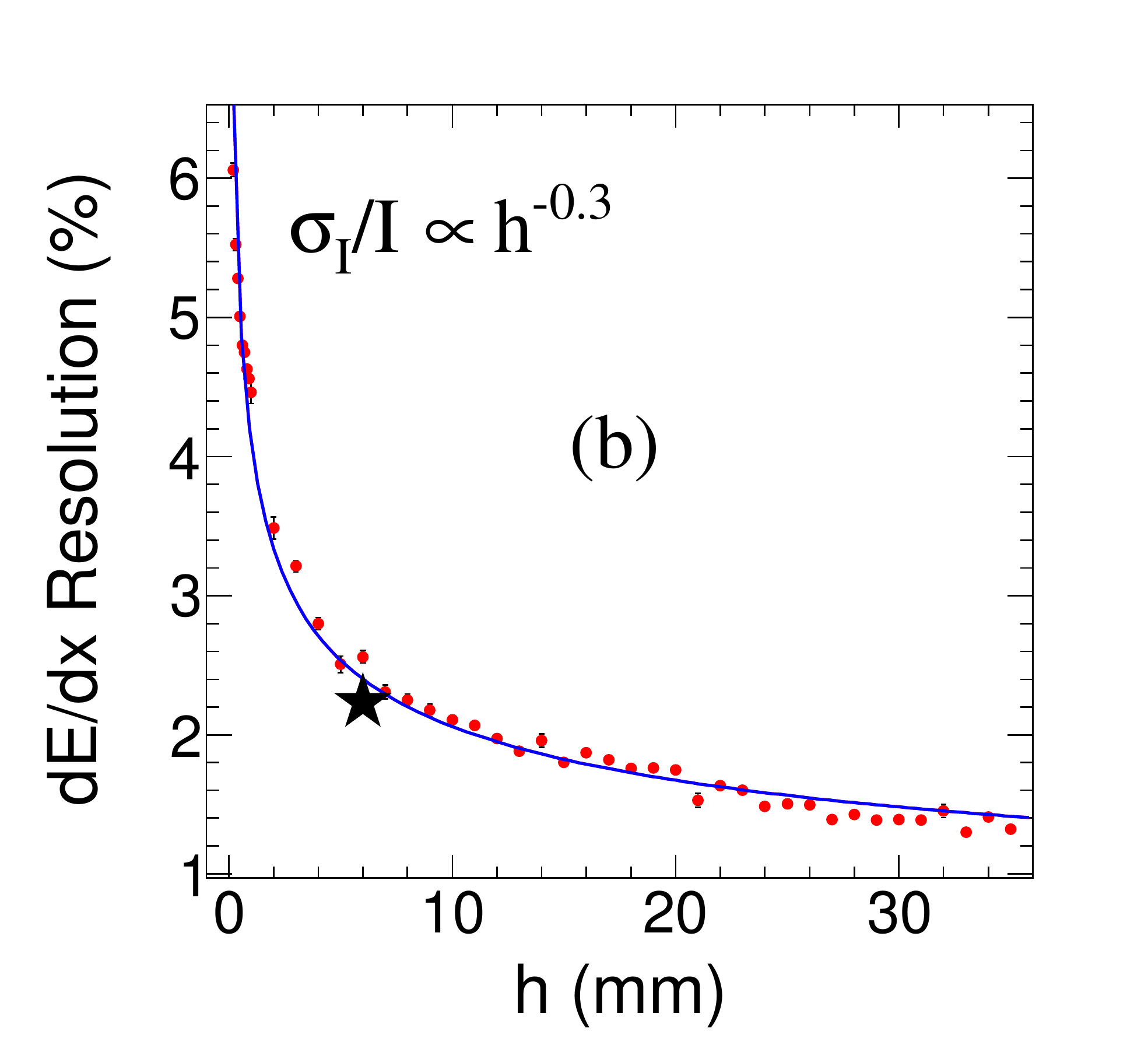}
    \hfill
  \includegraphics[width=0.32\linewidth]{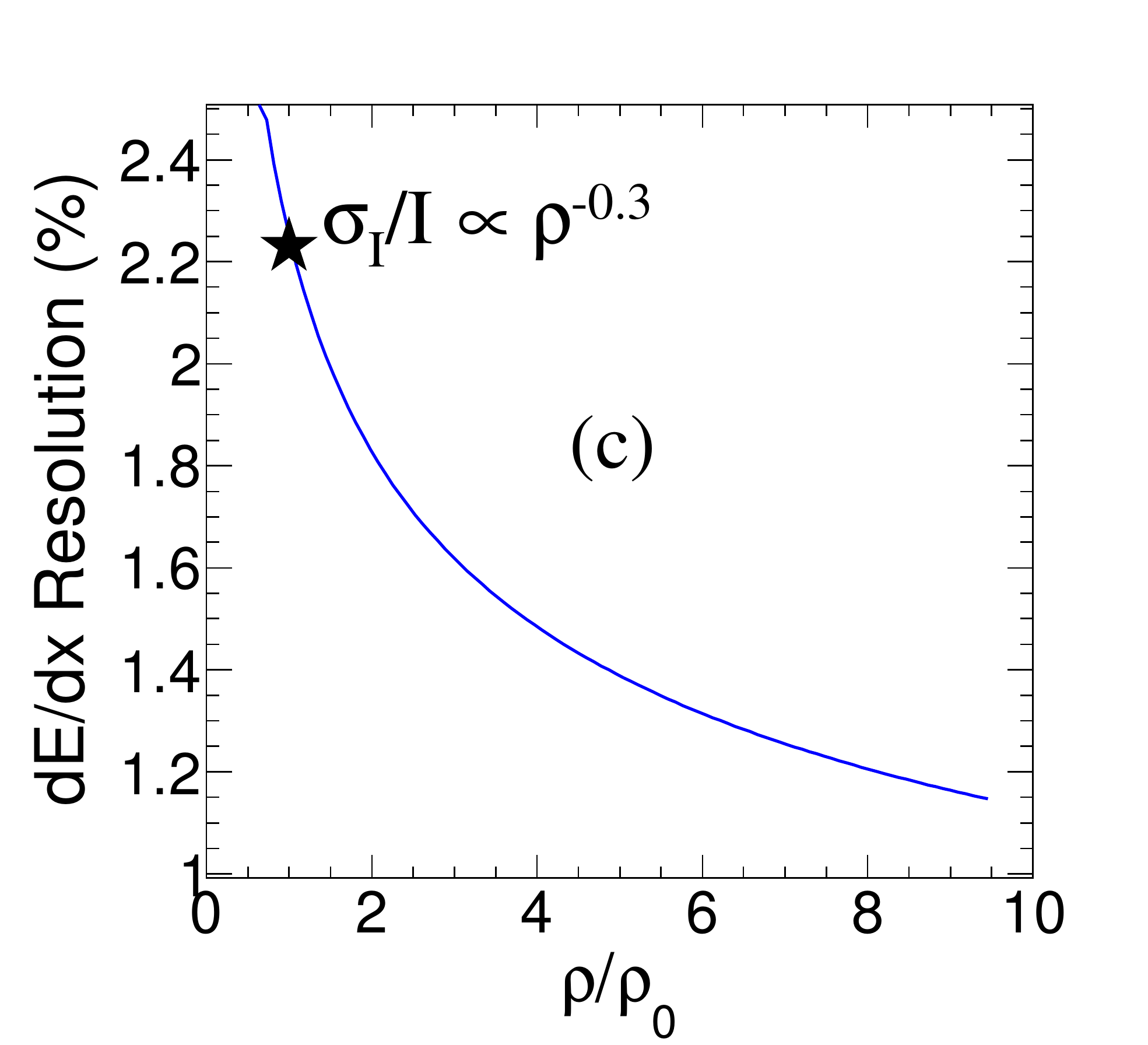}  \\
    \includegraphics[width=0.32\linewidth]{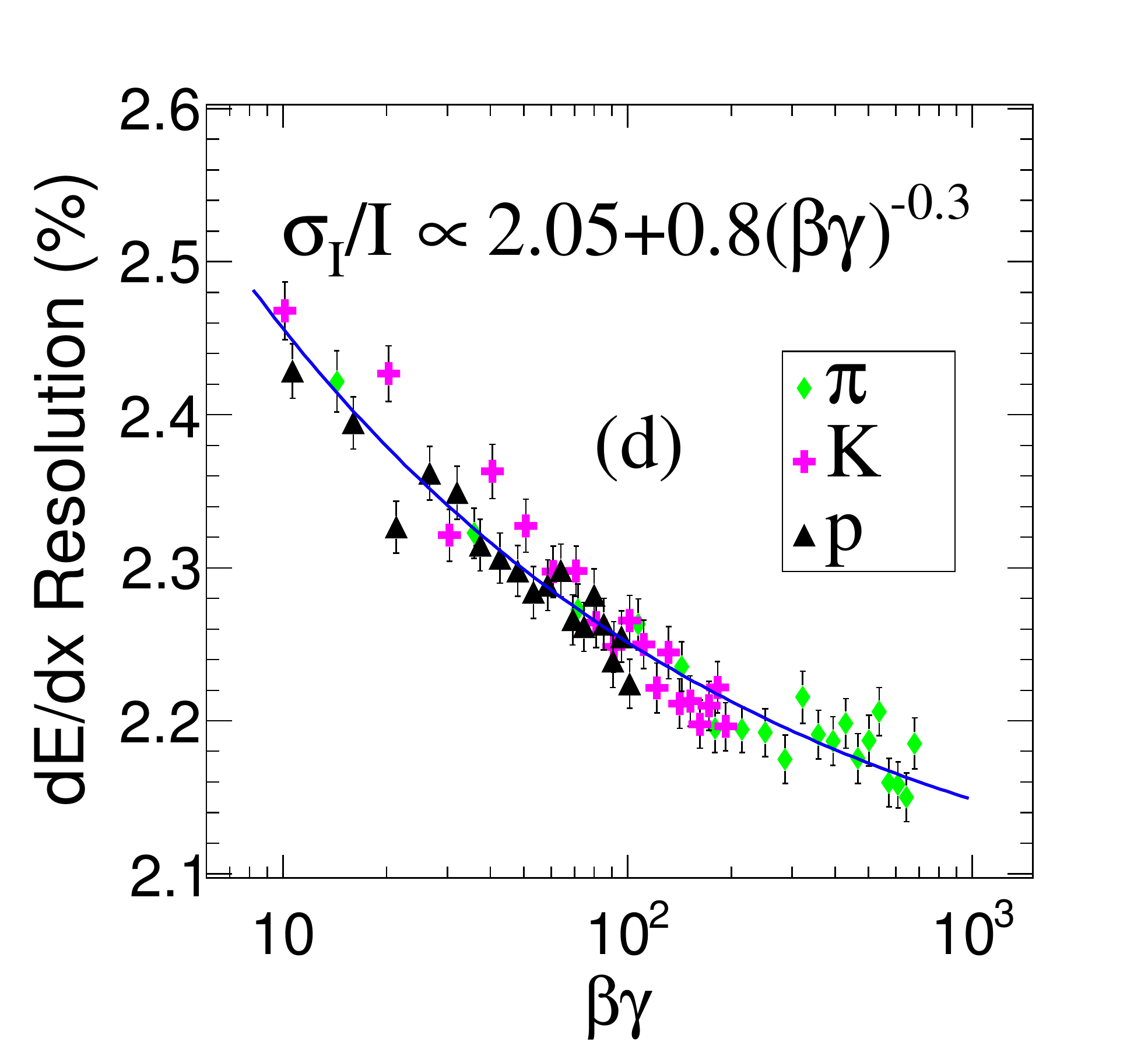}
    \hspace{0.01\linewidth}
  \includegraphics[width=0.32\linewidth]{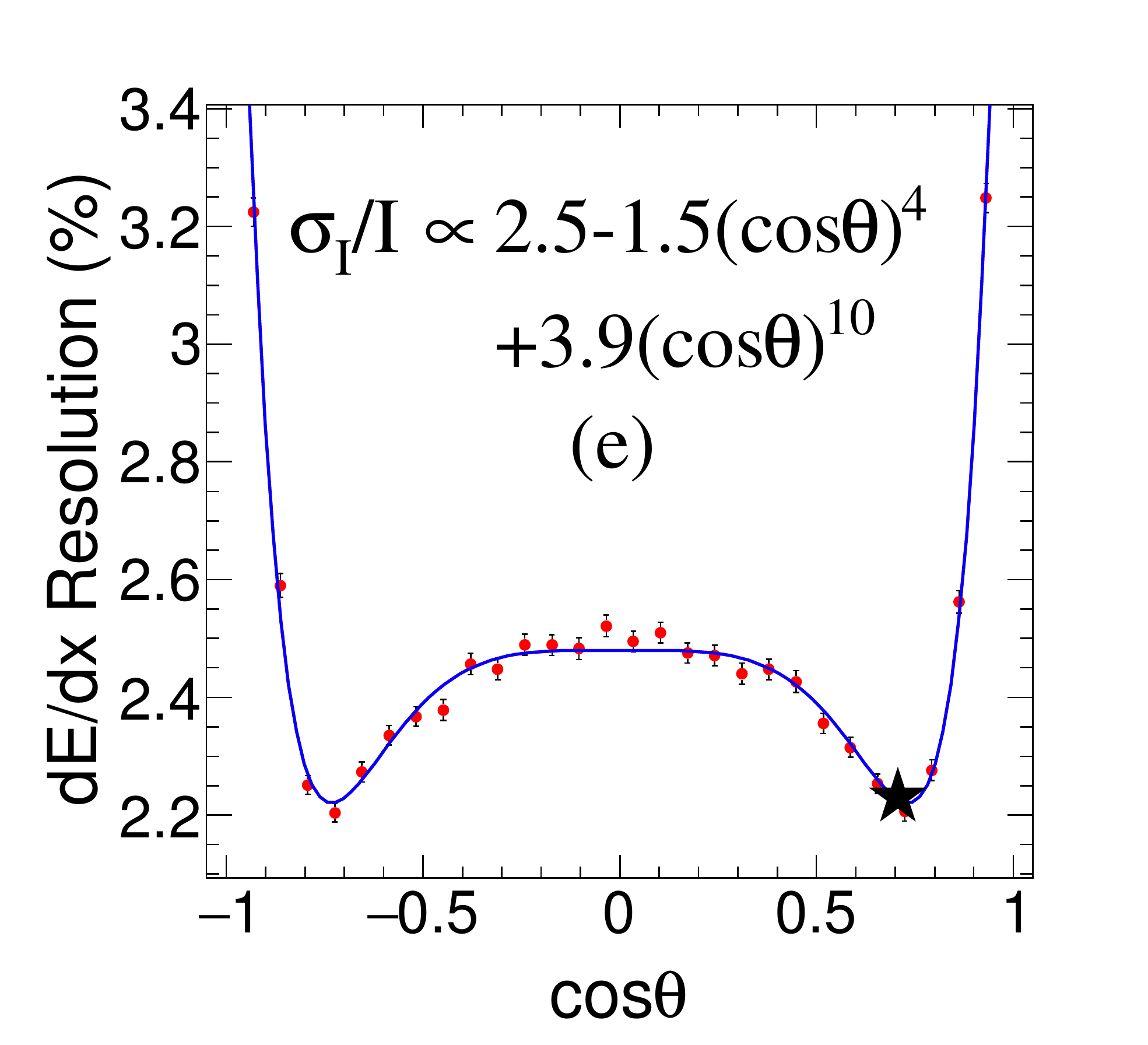}
  \caption{The intrinsic $dE/dx$ resolution versus the number of pad rings (a),
  the pad height along the radial direction (b), 
  the ratio of gas density $\rho$ over the default gas density $\rho_{0}$ 
  (equivalent to the ratio of corresponding pressures) (c),
  the relativistic velocity $\beta\gamma$ (e)  and $\cos\theta$ (f) of the ionizing particle.
  The default working point is indicated with a solid star symbol.
  Solid lines represent the fit projections.}
  \label{fig:dedx2var}
\end{figure*}

The intrinsic $dE/dx$ resolution arises  from  fluctuations at the primary ionization stage.
It depends on the number of the pad rings  $n$,  the pad height along the radial direction $h$, the density of the working gas $\rho$,
the relativistic velocity $\beta\gamma$ and the polar angle $\theta$ of the particle trajectory.
The resolution dependence   on these variables is 
studied using single-particle MC events.
We scan each variable to obtain its relationship with the intrinsic resolution.
Except for the variable under consideration, all others are kept constant
at their default values given in Sec.~\ref{sec:dedx_mean}, i.e., 
$n=222$, $h=6$ mm, $\rho=\rho_{0}=$1.73 mg/cm$^{3}$ and $\theta=45^{o}$
for pions with a momentum of 20 GeV/c.
The MC results are shown in Fig.~\ref{fig:dedx2var}.
 
The correlations between the variables are small. To a good approximation, the parameterization 
of $\sigma_{I}/I$  can be factorized as
% in Eq.~(\ref{eq:dedx_res}):

\begin{equation}
\begin{split}
\frac{\sigma_{I}}{I}  = &
\frac{13.5}{n^{0.5} \cdot (h\rho)^{0.3}}
[2.05+0.8(\beta\gamma)^{-0.3}]  \\
&\times [2.5-1.5(\cos\theta)^{4}+3.9(\cos\theta)^{10}],
\end{split}
  \label{eq:dedx_res}
\end{equation}
where $h$ and  $\rho$ are in 	 of mm and mg/cm$^{3}$, respectively.
%The normalization is scaled so
% that a 20 GeV/c pion  traversing the default working gas under standard 
%pressure and temperature with $\theta=45^{o}$,
%$n=222$ and $h=6$ mm has the  relative resolution of 2.25\% as in MC.
%By default, except the variable under investigation, others are fixed at the nominal values
%given in Sec.~\ref{sec:dedx_mean}.
% To check the correlation between the variables,
%we perform validation studies in multiple dimensions with $n$ ranging from 30 to 350,
%$h$ from 1 to 35 mm, $\rho$ from 0.16 to 20 mg/cm$^{3}$ for  argon- or helium-based gases,
%and $\beta\gamma$ from 6 to 1000. 
%When $\beta\gamma$>1000,  corresponding to the Fermi plateau
%in Fig.~\ref{fig:dedxMean2bg}, the resolution remains constant. 
 To check the correlation  between the variables, we validate the factorization  in the five-dimensional space by varying the 
variables within the ranges shown in Fig.~\ref{fig:dedx2var}.
In addition, the influence of the magnetic field is found to be negligible on the $dE/dx$ resolution.
When the magnetic field  is set to zero,
the induced relative change of  $\sigma_{I}/I$ is within 3\% for particles with momenta  larger than 1 GeV/c.

\begin{figure*}
  \centering
   \includegraphics[width=0.33\linewidth]{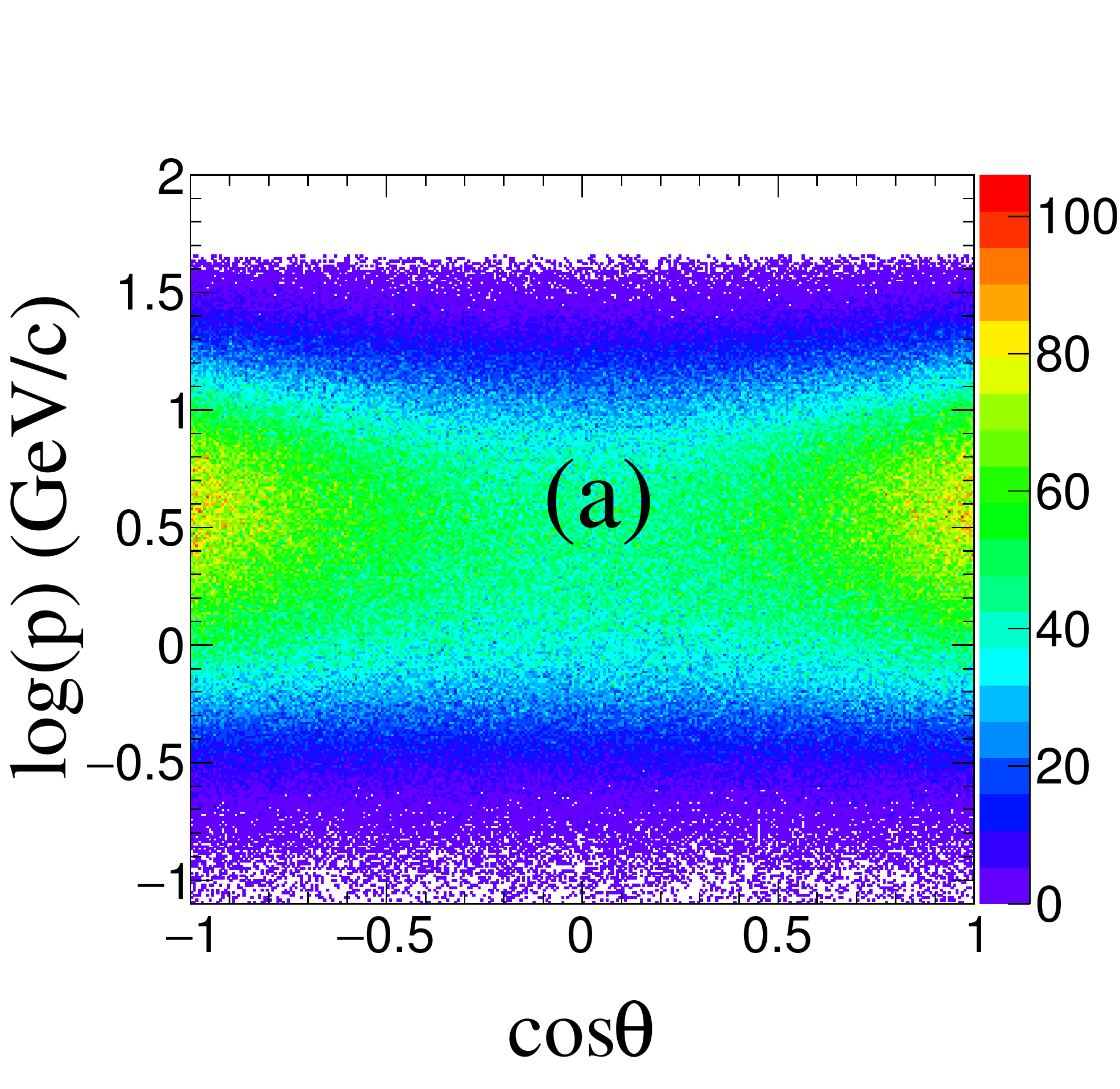} \hspace{2.5mm}
  \includegraphics[width=0.31\linewidth]{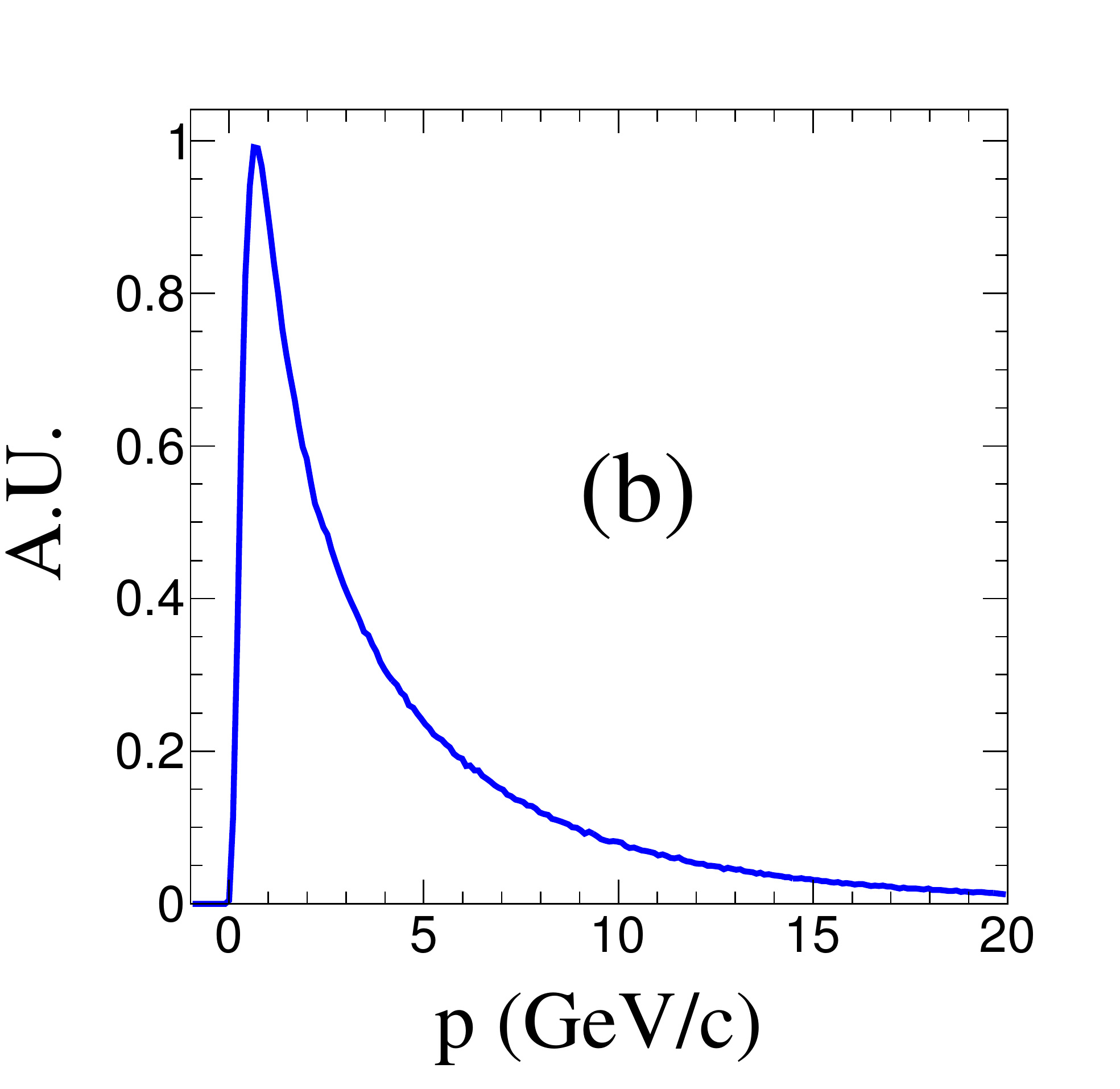} \hspace{1.mm}
  \includegraphics[width=0.31\linewidth]{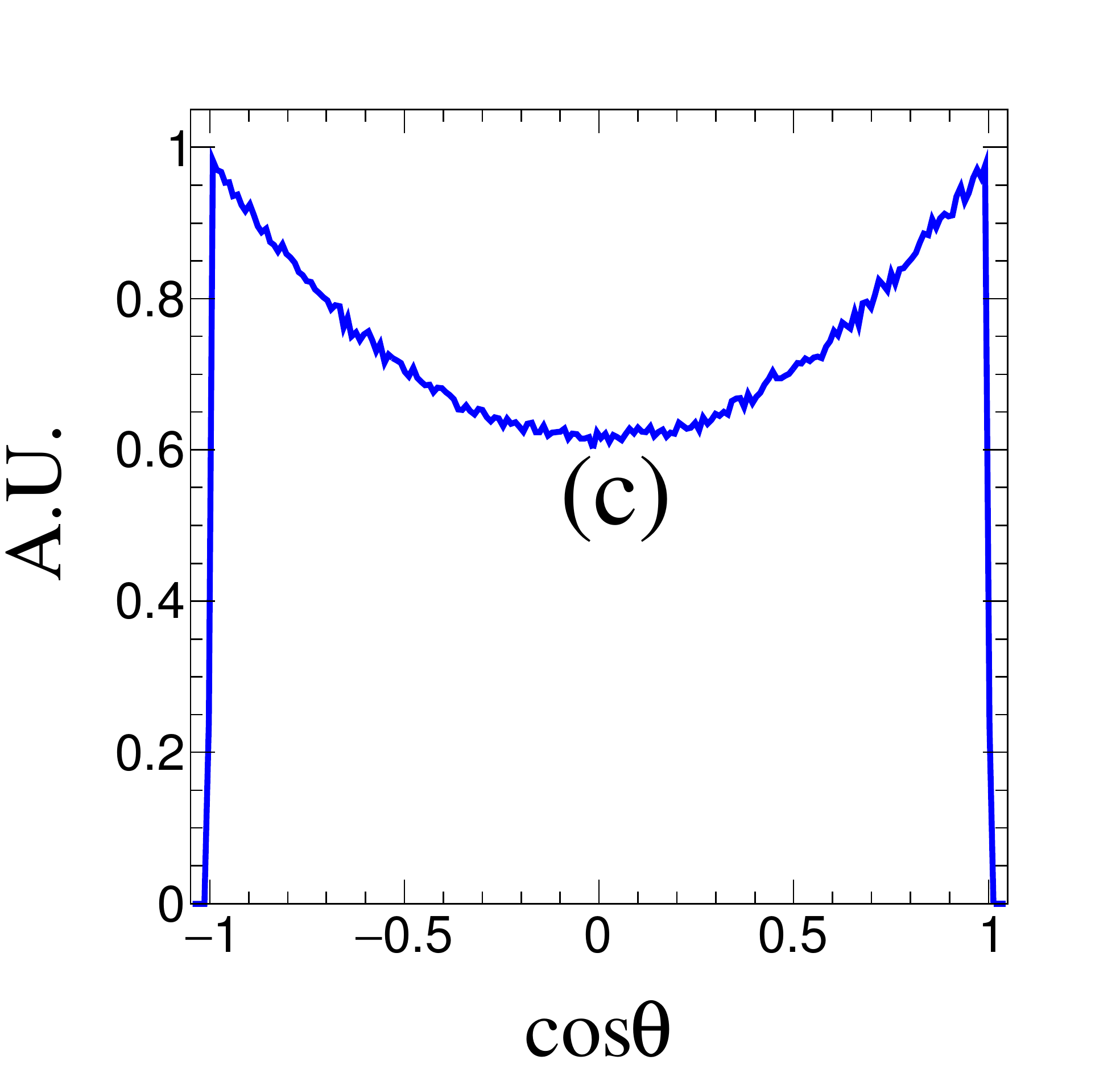}
 \caption{Kinematic distribution  of  kaons
in $e^+e^- \to Z \to q\bar{q}$ MC events as a function of 
 $\log(p)$ and $\cos\theta$ (a), 
 $p$ (b), and 
$\cos\theta$ (c).} 
  \label{fig:pcos_qq}
\end{figure*}

As  Eq.~(\ref{eq:dedx_res})  is derived from single-particle events,
its applicability   to physics events is validated using  kaons  from $e^+e^-\to Z\to q\bar{q}$ MC events.
The kinematic  distributions  are shown  in Fig.~\ref{fig:pcos_qq}.
We integrate Eq.~(\ref{eq:dedx_res})  over the $\cos\theta$ distribution given in Fig.~\ref{fig:pcos_qq}
and calculate the average $dE/dx$ resolution versus $\beta\gamma$. It is found to be consistent with the 
one directly obtained from MC. For example, for  kaons  with a momentum of 5 GeV/c in hadronic decays 
at the $Z$ pole, the intrinsic $dE/dx$ resolution is  3.1\%.

\subsection{Expected actual $dE/dx$ resolution of the CEPC TPC}
\label{sec:data_com}

In real experiments,  both detector effects and imperfect calibration can deteriorate the  $dE/dx$  resolution.
%In real experiments, deterioration of the resolution will arise
%due to the electronic limits  and
%imperfect calibration of track length calculation,  the working gas  fluctuation,
%hit overlap of close tracks, cross talk between the adjacent pads,
%the saturation effect and so on.
We  estimate the   potential degradation in previous TPCs by 
 comparing their experimental achievements 
 with the corresponding intrinsic $dE/dx$ resolutions obtained from MC simulation.
 
The TPCs considered in this study are summarized in Table~\ref{tab:com1}.
%Some other wire chambers are listed in the appendix for complementation.
The information about the experiments, unless explicitly stated, is taken from the references listed in the first row.
All the factors influencing the intrinsic resolution are implemented in  MC simulation,
including the composition of the working gas, the geometry of the TPC, the control samples and the truncation ratio 
used to remove the Landau tail. In the MC study, we resort to single-particle events 
and make them have identical particle type and  kinematic distributions 
with the corresponding control samples used in the experiments.
For the case where minimum ionizing pions are used,
 we assume a flat $\cos\theta$ distribution in the simulation when
  their $\cos\theta$ spectrum is not provided  in the references.
  The relative uncertainty arising from
such an approximation is estimated to be within a few percent and can be neglected.

%\begin{savenotes}
\begin{table*}
\begin{threeparttable}[htb]
 \caption{Properties of TPCs in previous experiments.
  Comparison of the relative  $dE/dx$  resolution between MC and experimental measurements.} 
\begin{tabular*}{\textwidth}{@{\extracolsep{\fill}}lccccccc@{}}
%\renewcommand\arraystretch{1.2}
% \fontsize{2.5mm}{3.4mm} 
 %\selectfont
 %\centering
%\begin{tabular*}{1\linewidth}{@{}l|c|c|c|c|c|c|c}  
\hline

Experiment    &  PEP-4        &  TOPAZ    &   DELPHI   &   ALEPH     &      STAR     &    ALICE    & CEPC \\
          & \cite{pep1,pep2,pep3} &  \cite{topaz}  & \cite{delphi1,delphi2}  &  \cite{aleph1,aleph2} 
          & \cite{star1,star2}   &    \cite{alice1,alice2}    &  \\

\hline
Start of data taking       &   1982        &  1987     &      1989   &   1989      &     2000      &  2009     & ---\\

Colliding Particles  &  e$^{-}$/e$^{+}$   &  e$^{-}$/e$^{+}$   &  e$^{-}$/e$^{+}$  &  e$^{-}$/e$^{+}$  &  Au/Au  &  p/p   &  e$^{-}$/e$^{+}$  \\

$E_{\rm{beam}}$ (GeV)   &   14.5   &    26   &   45.6  &    45.6  &   100  &   1380   &   125  \\

\hline

          &  Ar: 0.8    &   Ar: 0.9    &  Ar: 0.8      & Ar: 0.91    &   Ar: 0.9   & Ne: 0.857  &       Ar: 0.93 \\
  Gas           &  CH$_{4}$: 0.2   &   CH$_{4}$: 0.1   &  CH$_{4}$: 0.2     & CH$_{4}$: 0.09    &   CH$_{4}$: 0.1  & CO$_{2}$: 0.095   &   CH$_{4}$: 0.05 \\
             &               &                &                 &              &          & N$_{2}$: 0.048        &  CO$_{2}$: 0.02 \\

Pressure (atm) &  8.5        &  3.5           & 1              &  1            &  1            &  1    & 1  \\
$\rho$ (mg/ml) &  12.43      &  5.47          & 1.46           &  1.57         &  1.56        &  0.95  & 1.73  \\	
$n$              &  183        &  175           &  192           &  344          &  13, 32 \tnote{3}   & 63,64,32  \tnote{3}  &    222 \\
%\hline
$h$ (mm)         &  4          &  4             &  4             &  4           &  12, 20  \tnote{3}      & 7.5,10,15  \tnote{3}  &    6   \\
%\hline
Length (mm)       & 2000      &     3000    &    2680      &       4400     &     4200    &   5000     &   4700  \\      
%\hline
Control Sample   &  $e$          &  $\pi$         &  $\pi$         &  $e$           &  $\pi$      &  $\pi$    &    $K$  \\
$p$ (GeV/c)          &  14.5         &  0.4-0.6       &  0.4-0.6       &  45.6        &    0.4-0.6          &   6.0        &   5.0  \\
%\hline
Truncation range  & 0-65\%      & 	0-65\%        &  8-80\%        & 8-60\%     & 0-70\%     & 0-60\%    &   0-90\% \\
%\hline  
%$(\sigma_{I}/I$)$_{\theta=45^{o}}$  &  2\%  &  2.6\%  & 3.4\% & 2.3\%	  &  {\red 4.2\%}  &  {\red 3.2\% }   &   {\red 2.3\%}  \\ 

 $N_{\rm{eff}}$      & $n$           &  0.7$n$ \tnote{1}    &  0.6$n$ \tnote{2}       & 338                &    44        &    149      &     $n$ \\
%\hline
\hline
$(\sigma_{I}/I)_{\mathrm{MC}}$     & 2.6\%   &  3.8\%    &  5.4\%   &   3.0\%   &   5.3\%  &     3.3\%  &    3.1\%   \\

$(\sigma_{I}/I)_{\mathrm{exp}}$    & 3.5\%  &  4.6\%    &  6.2\%   &   4.4\%   &   6.8\% \tnote{4}
&    5.0\%    &    ---   \\
%\hline
$\left| \frac{(\sigma_{I}/I)_{\mathrm{exp}}}{(\sigma_{I}/I)_{\mathrm{MC}}} -1\right |$                            &  0.35    &  0.21    &  0.15    &  0.47     &   0.28   &    0.52    &    ---  \\
%\hline
%$\sqrt{\sigma_{exp}^{2}-\sigma_{MC}^{2}}$  &  2.3\%  &   2.6\%   &   3.0\%    &   3.2\%   &     4.3\%   &   3.8\%     &       \\
%\hline
%$\delta G$ (\%) &         &   0.9     &      &  3.0    &      &    1.5       &     \\
%\hline
\hline
\end{tabular*}
  \label{tab:com1}
  
   \begin{tablenotes}
            \item [1] It's  required that at least  78 hits are used for the $dE/dx$  calculation~\cite{topaz}. 
Here we assume there are 70\% effective hits.
            \item [2] 65\% tracks have more than 40 isolated hits, and 35\% more than 100~\cite{delphi1}.
We assume there are 60\% effective hits.
            \item [3] In the STAR and ALICE detectors, the inner, intermediate and outer subdetectors have different pad sizes. 
            \item [4] See Fig. 21 in Ref.~\cite{star2}.
          \end{tablenotes}
  \end{threeparttable}
\end{table*}

 Besides the factors discussed above,
the  number of the effective hits used for the  $dE/dx$  calculation, denoted as $N_{\rm{eff}}$, 
is also considered because it  greatly influences the $dE/dx$ resolution in  the earlier experiments.
In TOPAZ and DELPHI, for example,
on average only 60-70\% effective TPC hits are available for tracks
in jets due to the large size of their  TPC readouts and resulting in serious hit overlap.
 STAR and ALICE have made significant progress in exploiting   high-granularity readouts to handle their dense tracking environment.
% But in heavy ion collisions the event multiplicity is quite high with thousands of tracks,
%the effective hits in jets can only account for 70\% of the total, as is shown in the STAR experiment.
In ALICE, the fraction of $N_{\rm{eff}}$ is about 93\% or even larger in proton-proton collisions~\cite{alice2}.
Compared to ALICE, the  CEPC TPC will have a higher granularity and endure much smaller track multiplicities.
Therefore we neglect this effect at the CEPC.
Even if assuming that  5\% of the hits are discarded,
the resulting relative change in the $dE/dx$ resolution is within 3\% according to Eq.~(\ref{eq:dedx_res}).

In the last  row of Table~\ref{tab:com1},
the relative difference between the intrinsic and actual $dE/dx$ resolutions are listed.
It varies from  0.15 to 0.50 between the different experiments.
Studies performed by the ALICE TPC Collaboration \cite{alice3,alice4}
show that the main  detector effects  causing the deterioration include 
diffusion in the drift volume, fluctuations in the amplification and DAQ processes,
and cross talk between the readout pads. 
Based on MC simulation, we estimate that these effects will cause a degradation of at least  20\% at the CEPC TPC.
Therefore, we define two scenarios  for further discussion about the $dE/dx$ performance 
that might eventually be achieved by the CEPC,
namely an ``optimistic scenario'' and a ``conservative scenario'',
corresponding to degradations of 20\% and 50\%, respectively, with respect to the intrinsic $dE/dx$ resolution.

%Before any proofs of improvements in these  aspects,
%we conservatively assume 0.5 deterioration at the CEPC TPC.
% This gives us  an  $dE/dx$ resolution approximately of  4.6\%  for pions in jets with a momentum of 5 GeV. 
%This estimate is close to the expected $dE/dx$ resolution of the TPC at the 
%ILC, which  has a similar design as the CEPC. 
%A resolution of 4.7\% is predicted for the ILC TPC based on their studies with prototypes\cite{ilc_proto}.
%
%In a performance study of a GEM-based readout  TPC module using electron beam,
% they estimate that the relative  $dE/dx$  resolution of the ILC TPC would be around 4.7\% 
% for 5 GeV/c electrons on the Fermi plateau~\cite{ilc_proto}.}

\subsection{ Expected PID performance of the CEPC TPC}

\begin{figure}
  \centering
  \includegraphics[width=0.9\linewidth]{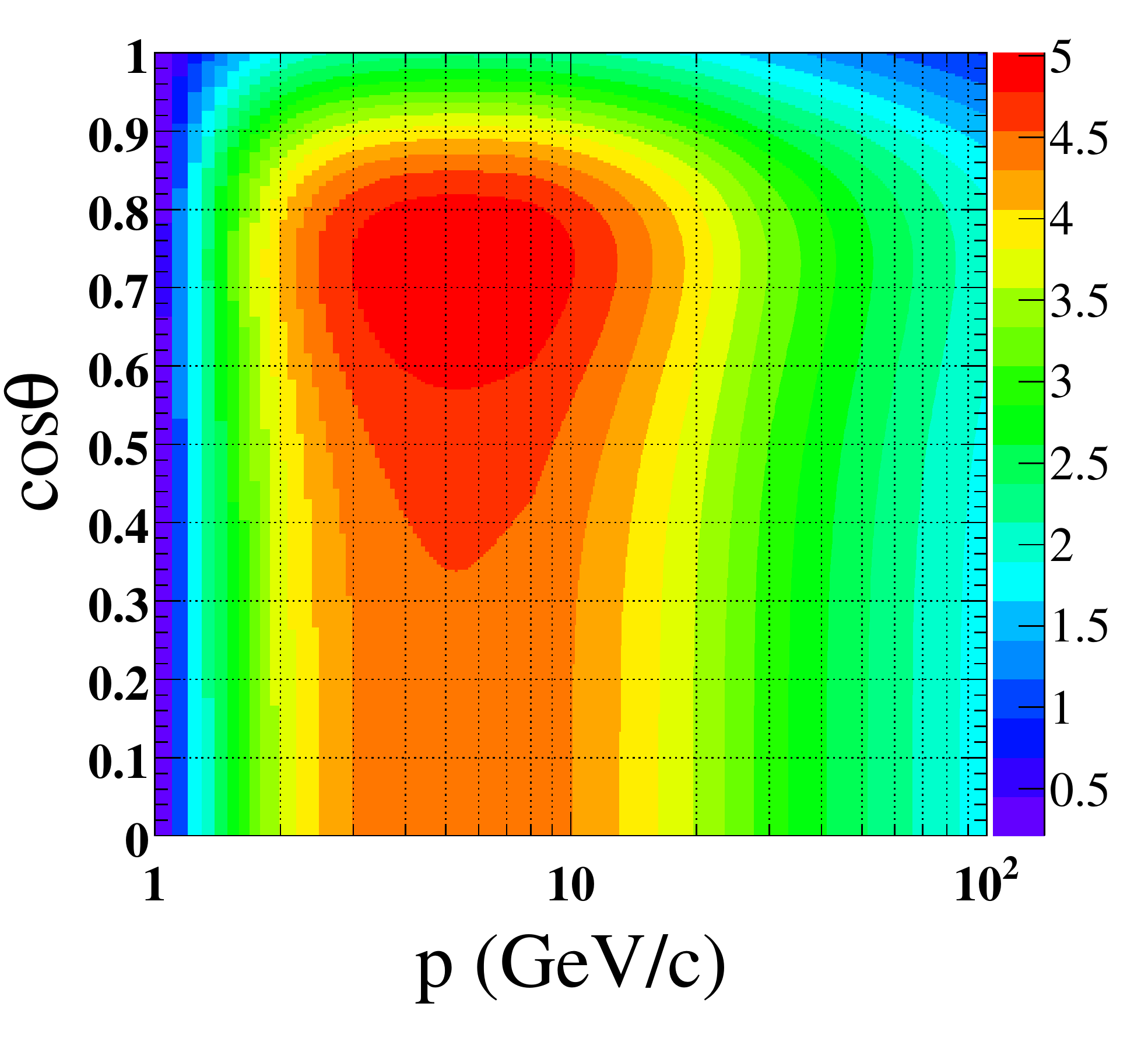}
  \caption{Separation power $S_{K\pi}$ between kaons and pions  in the $p$-$\cos\theta$ plane
using $dE/dx$ measurements of the CEPC TPC for the ideal simulation.}
  \label{fig:sep2p2cos}
\end{figure}

A common figure of merit for the PID performance is the separation power $S$.
Between particle types $A$ and $B$ we define

\begin{equation}
S_{AB} = \frac{|I_{A}-I_{B}|}{\sqrt{{\sigma^{2}_{I_{A}}}+\sigma^{2}_{I_{B}}}},
%S_{AB} = \frac{|I_{A}-I_{B}|}{\sqrt{\sigma_{I}_^{2}+\sigma_{IB}^{2}}}.
  \label{eq:sep}
\end{equation}
\noindent
where $I_A$ ($I_B$) and $\sigma_{I_{A}}$ ($\sigma_{I_{B}}$)
are the average $dE/dx$ measurement of particle type A (B) and the corresponding resolution. 
 In the ideal case assuming no degradation and $\sigma_{I}$ follows Eq.~(\ref{eq:dedx_res}),
we estimate  $S_{K\pi}$ at the CEPC as a function of $p$ and $\cos\theta$
(see Fig.~\ref{fig:sep2p2cos}).

\begin{figure*}
  \centering
  \includegraphics[width=0.32\linewidth]{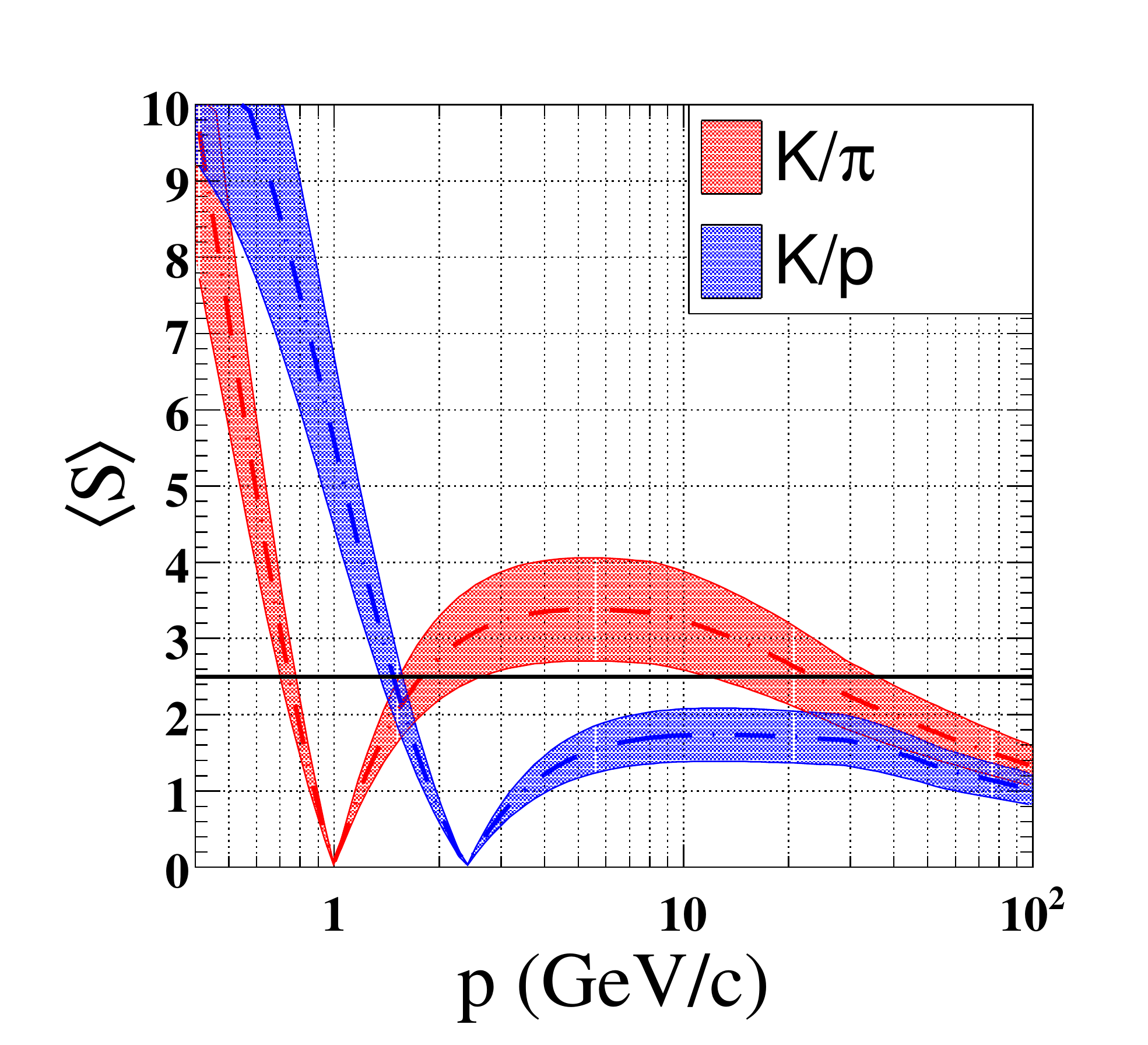}
  \hfill
    \includegraphics[width=0.32\linewidth]{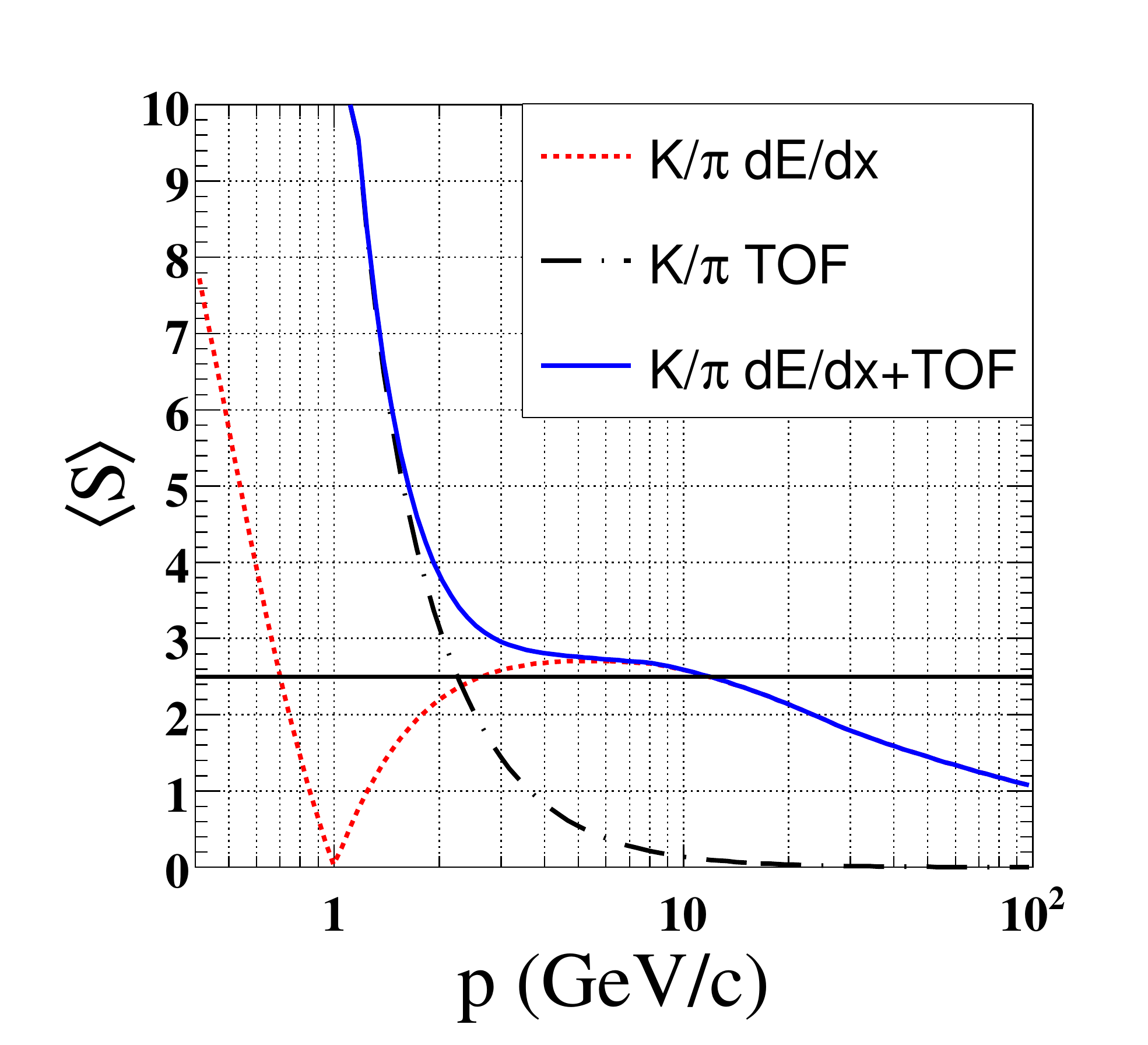}
    \hfill
    \includegraphics[width=0.32\linewidth]{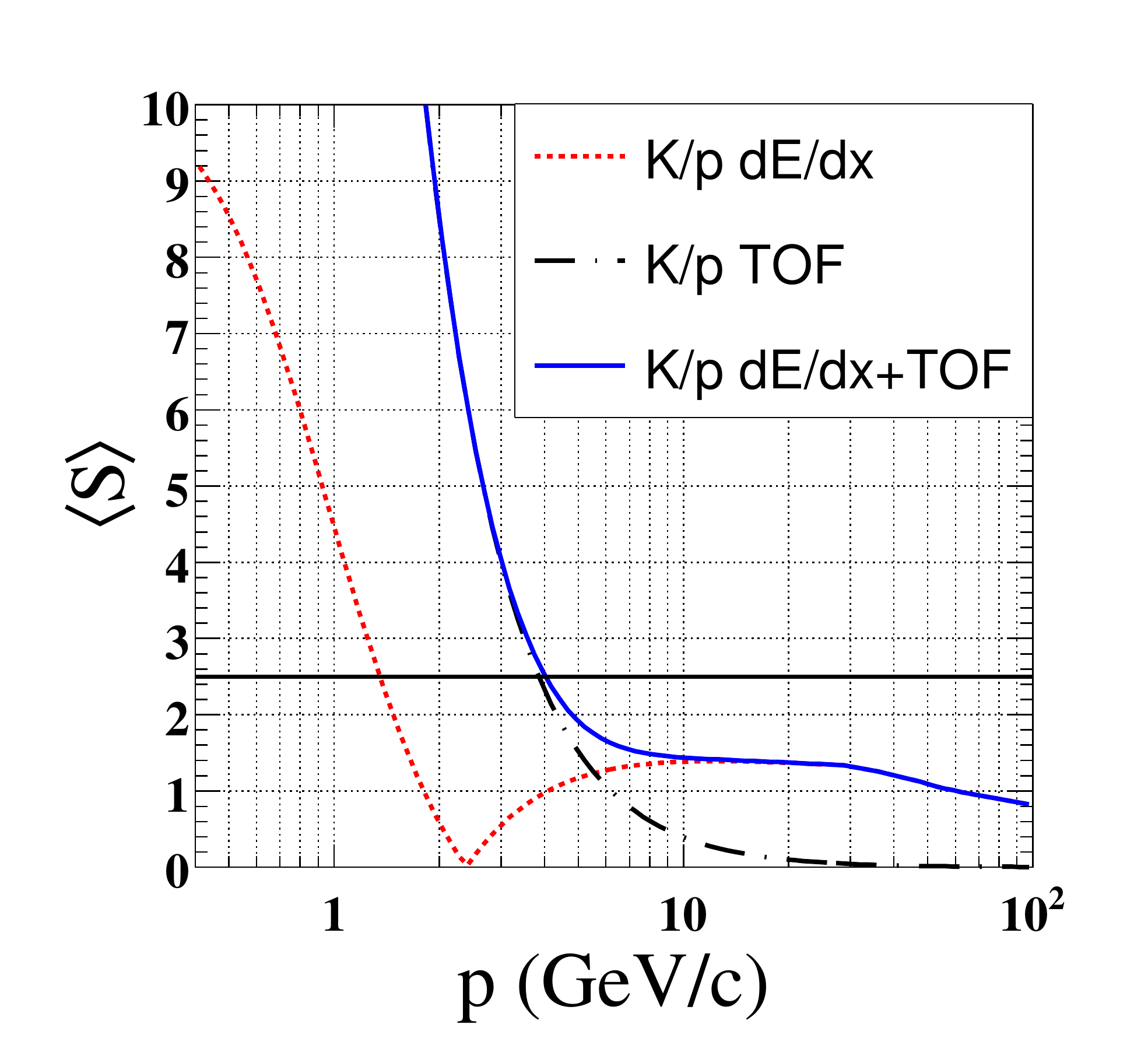}  
  \caption{Average separation power  $\big< S\big>$ versus momentum between different particle types in hadronic decays at the $Z$ pole.
  Left: only  $dE/dx$ is used. The bands delimit the area
  between the ideal simulation and the conservative scenario for the CEPC TPC.
  The optimistic scenario is shown as dash-dotted lines.
  Middle and right:  $dE/dx$ (in the conservative scenario) and/or TOF are used for $K/\pi$ and $K/p$ separation.
  The black solid line corresponds to 2.5 $\sigma$ separation.}
  \label{fig:sep2p}
\end{figure*}

One often cares about the average separation power $\big< S \big>$ versus momentum
after integrating over the $\cos\theta$ dimension.
Given the $\cos\theta$ distribution in $e^+e^-\to Z\to q\bar{q}$  decays (see Fig.~\ref{fig:pcos_qq}),
the plots of $\big< S_{K\pi}\big>$  and  $\big< S_{Kp}\big>$ as a function of $p$ are shown in Fig.~\ref{fig:sep2p}.
In the left plot, the separation powers using $dE/dx$ for different TPC performance scenarios are illustrated.
One can see that $dE/dx$  alone is incapable of $K/\pi$ separation around 1 GeV/c
and yields poor $K/p$ separation beyond 1.5 GeV/c.
To overcome this disadvantage, the exploitation of TOF information is considered.

According to a recent study on the CMS high-granularity calorimeter \cite{cms_ecal},
precise TOF information could be provided by the CEPC ECAL with a precision  of tens of picoseconds.
Supposing TOF information with a 50~ps time resolution, and given the $dE/dx$ measurements in the conservative scenario, 
the average $K/\pi$ and $K/p$  separation powers  are calculated using both $dE/dx$ and TOF.
They are shown in the middle and right plots of Fig.~\ref{fig:sep2p}.
%Under the magnetic field of 3 T,  the flight path averaged over all directions of particles in the  $e^+e^-\to Z\to q\bar{q}$  sample is larger than 2.4 m for particle momenta above {\bf XXX}.
Accounting for the time resolution and the location of the ECAL, the TOF information can provide $K/\pi$ ($K/p$) 
separation better than 2.5 $\sigma$ up to 2.1 (4.0) GeV/c. By combining TOF and $dE/dx$, 
more than 2.0 (1.4) $\sigma$ $K/\pi$ ($K/p$) separation can be achieved up to 20 GeV/c.

The PID performance depends on the kinematic distributions and relative abundance of the 
charged particles in the sample under study. As an example, we take the process $e^+e^-\to Z\to q\bar{q}$ 
(see Fig.~\ref{fig:pcos_qq}) with an average of 20 charged particles per event, of which  85\% are 
pions, 10\% are kaons, and 4\% are protons.  
We calculate the  average separation powers $\big< S_{K\pi}\big>$ and $\big< S_{Kp}\big>$ 
for particles with momenta in the range from 2-20 GeV/c.
They are listed in Table~\ref{tab:sep}.
Particles with momenta smaller than 2 GeV/c are not considered since 
they can be clearly separated.

Due to the importance of the kaon selection performance for flavor physics,
we also provide an estimation of the 
kaon selection efficiency $\varepsilon_{K}$ and the corresponding  purity $p_{K}$,
together with the probability of mis-identifying pions (protons) as kaons  $p_{\pi (p)\to K}$.
They are defined as 
\begin{equation}
\begin{split}
&\varepsilon_{K} =  \frac{N_{K\to K}}{N_{K}}, \\
& p_{K}  = \frac{N_{K\to K}}{N_{K\to K}+N_{\pi\to K}+N_{p\to K}},  \\
& p_{\pi\to K} = \frac{N_{\pi\to K}}{N_{\pi}}, \\
&p_{p\to K} = \frac{N_{p\to K}}{N_{p}},
 \end{split}
  \label{eq:misID}
\end{equation}

\noindent
where $N_{K}$, $N_{\pi}$, $N_{p}$ are the total numbers of generated kaons, pions and protons that traverse
the innermost pad ring of the TPC, $N_{K\to K}$ is the number of  correctly identified kaons,
and $N_{\pi(p)\to K}$ is the number of pions (protons) mistakenly identified as kaons. 

\begin{figure}[htp]
  \centering
    \includegraphics[width=0.9\linewidth]{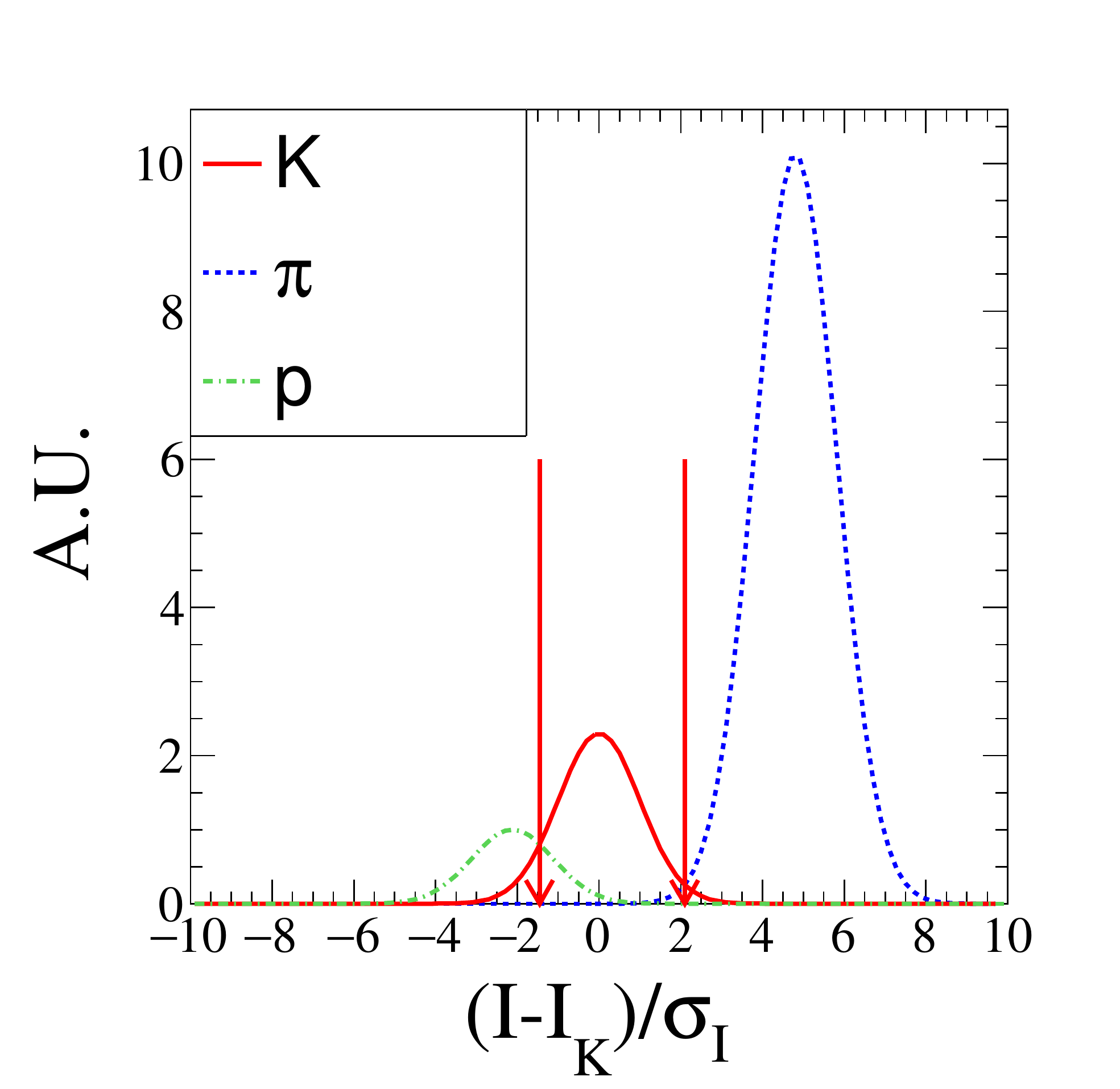}
  \caption{The scaled spectra of $(I-I_{K})/\sigma_{I}$ using $dE/dx$ measurements alone
  for particles with a momentum of 5 GeV/c,  assuming a 20\% degradation.
  The relative populations are $N_{\pi}=4.4 N_{K}$ and $N_{K}=2.3 N_{p}$ according to MC simulation.}
  The intersections marked by the arrows are chosen as the cut points.
  \label{fig:kpip_dedx}
\end{figure}

The kaon selection  is performed based on the variable  $(I-I_{K})/\sigma_{I}$,
where  $I$ and $I_{K}$ are the experimental measurement (either by $dE/dx$ alone or  by combining $dE/dx$ and TOF)
 and the expected value for the kaon hypothesis respectively,
and $\sigma_{I}$ denotes the experimental resolution.
Their spectra should be close to Gaussian distributions with a width of 1.
In Fig.~\ref{fig:kpip_dedx} we illustrate the scaled spectra of kaons, pions and protons with a momentum of 5 GeV/c 
using $dE/dx$  alone assuming  a 20\% degradation.
According to Eq.~(\ref{eq:sep}),
the peaks between the kaon and pion (proton) spectra should be $\sqrt{2}S_{K\pi}$ ( $\sqrt{2}S_{Kp}$) apart,
where $S_{K\pi}$ ( $S_{Kp}$) is the corresponding separation power.
 The relative populations $N_{\pi}/N_{K}$ and $N_{K}/N_{p}$ 
vary versus momentum and are determined based on MC simulation.
We choose the intersections of the spectra as the cut points  (marked by the arrows in the plot),
in order to calculate the kaon identification efficiency and purity together with the mis-identification rates according to Eq.~(\ref{eq:misID}).
We calculate these parameters at each momentum point from 2-20 GeV/c in  $e^+e^- \to Z \to q\bar{q}$ events (see Fig.~\ref{fig:pcos_qq})
and provide in Table~\ref{tab:sep} the average values.
 The MC sample under study is large enough ($\sim 8$  million) and the statistical errors are
negligibly less 0.1\%.

%as illustrated in Fig.~\ref{fig:kpip_dedx}. We integrate the distribututions In x GeV/c wide slices 
%of momentum  and use as integration boundaries the intersection points between the kaon and pion spectra, 
%and the kaon and proton spectra as integration boundaries to calculate the performance parameters.
%This integration boundaries depend on the relative population between the charged particle types, 
%The ratios given in the 
%table are the average values for particles with momenta in the range from 2-20 GeV/c.

\begin{table}[htp]
\centering
\caption{Expected PID performance parameters at the CEPC in different scenarios.
Shown are the average value of particles with momenta from 2-20 GeV/c
 in the  $e^+e^-\to Z\to q\bar{q}$  decays.}
\begin{tabular*}{\columnwidth}{@{\extracolsep{\fill}}c|cccc@{}}
%\renewcommand\arraystretch{1.2}
 %\fontsize{3.2mm}{3.5mm} 
 %\selectfont  
%\begin{tabular*}{0.73\linewidth}{c|c|c|c|c|c}  
\hline
               &  Deterioration        &       0       &   0.5        &   0.2            \\
          %     &   Total number of pads  &    $n_{\rm {pad}}$    &     $n_{\rm {pad}}$          &   $n_{\rm {pad}}$                  \\     
          %   &    Effective radius          &     $r$               &   $r$               &       $r$                     \\
                                \hline
      &  $\big< S_{K\pi} \big>$                                   &  3.9       &    2.6      &   3.2     \\
      &  $\big< S_{Kp}   \big>$                                      & 1.5      &   1.0         &   1.2       \\
  $dE/dx$           &   $\varepsilon_{K}$  (\%)             &    93.2        &     84.5       &   90.9        \\
       &  $p_{K}$   (\%)                             &  86.5      &   76.1       &   82.4  \\
           &    $p_{\pi\to K}$   (\%)              &    0.1       &      1.3         &     0.5      \\
      &     $p_{p\to K}$   (\%)                 &    33.0      &      47.2       &    40.1    \\
    
    \hline    
                   &  $\big< S_{K\pi} \big>$              &   4.0               &     2.8      &   3.4         \\
        $dE/dx$                       &  $\big< S_{Kp} \big>$              &    3.2                &   2.8      &  3.0         \\
         \&        &  $\varepsilon_{K}$    (\%)          &    96.8         &    90.4      &     95.0     \\
     TOF      &  $p_{K}$      (\%)                           &    97.0        &     90.1    &    94.5      \\
                    &    $p_{\pi\to K}$  (\%)     &     0.1          &    1.1       &     0.4     \\
              &     $p_{p\to K}$   (\%)                 &    6.4          &      13.8.  &      9.6        \\\hline
\end{tabular*}
  \label{tab:sep}
\end{table}

In  the ideal simulation, the $dE/dx$ measurements  ultimately provide roughly 4 $\sigma$ (1.5  $\sigma$)
separation between kaon and pion (proton) in inclusive $e^+e^-\to Z\to q\bar{q}$  decays.
The overall kaon identification efficiency reaches 93.2\% with a purity of 86.5\%.
 The PID performance is limited by the proton contamination.
By combining the $dE/dx$ and TOF measurements, the $K/p$ separation is greatly enhanced from 1.5 $\sigma$ to 3.2 $\sigma$.
As a consequence, the kaon identification efficiency is improved to 96.8\% with a corresponding purity of 97.0\% 
%and the proton contamination of the selected kaon sample is significantly suppressed.

In the conservative scenario, the kaon identification efficiency and purity degrade significantly 
%to 81.9\% and 75.5\%, respectively. 
 mainly due to the more serious proton contamination.
In this case, the TOF measurement plays a crucial role and can ameliorate the performance back to an efficiency of 90.4\% and 
a purity of 90.1\%. 
If the optimistic scenario can be realized at the CEPC, by combining $dE/dx$ and TOF, 
we expect the efficiency  reaches 95.0\% for  kaon identification with a purity of 94.5\%, which is only slightly 
degraded from the ideal simulation.
In all scenarios,  the pion mis-identification rate can  be controlled at a 1\% level.

\section{Conclusion}
\label{sec:conclusion}

Effective particle identification will enrich the CEPC physics program, 
especially when operating at the $Z$ pole. 
Using a GEANT4-based MC simulation, we study the PID performance at the CEPC based on
the  $dE/dx$  measurements in the TPC  and 
 the TOF information provided by the ECAL with an
assumption of 50 ps time resolution.

 We  explore the  kaon identification performance in the momentum range from 2-20 GeV/c
 in inclusive hadronic $Z$ decays,
showing that an effective kaon identification can be achieved with the combined 
information of $dE/dx$ and TOF.
If the degradation of the $dE/dx$ measurements due to 
detector effects can be controlled to less than 20\%,
both the average kaon identification efficiency and purity can approach 95\%.
%It is shown that by combining  $dE/dx$  and the
%timing information from the ECAL, 
 %an effective kaon identification 
%with around 90\%  efficiency and purity 
%at the $Z$ pole operation
%can be achieved even with the most conservative detector assumption.
%On top of the $dE/dx$ measurements, 
%the TOF provides essential complementary information for the kaon identification.
%On the one hand, the TOF  remedies the $K/\pi$ separation around 1 GeV/c where 
%the $dE/dx$ separation ability vanishes.
%And on the other,  it makes significant impact on the $K/p$ separation over the objective momentum range (2-20 GeV/c). 
%An improved PID performance is feasible if the degradation of the $dE/dx$ measurements due to 
%detector effects can be controlled to less than 20\%. 
 More detailed microscopic simulation and beam tests are expected to validate these conclusions in the future.

\begin{acknowledgements}
We express our great appreciation to
the technical staff in our institutions for their  discussion. 
In particular, we would like to thank Prof. Peter Christiansen of  Lund University
for his very helpful suggestions.

This work was supported by  
National Key Program for S\&T Research and Development under Contract Number 2016YFA0400400;
the Hundred Talent programs of Chinese Academy of Science under Contract Number Y3515540U1;
National Natural Science Foundation of China under Contract Number 11675202;
IHEP Innovation Grant under Contract Number Y4545170Y2;
Chinese Academy of Science Focused Science Grant  under Contract Number QYZDY-SSW-SLH002;
Chinese Academy of Science Special Grant for Large Scientific Project under Contract Number 113111KYSB20170005;
National 1000 Talents Program of China.

\end{acknowledgements}

\end{document}